\documentclass[aps,prl,nofootinbib,preprintnumbers,floatfix,twocolumn,superscriptaddress]{revtex4}
\usepackage{pgfplots}
\usepackage{graphicx}
\usepackage{amsmath}
\usepackage{bm}
\usepackage{latexsym}
\usepackage{ulem}
\usepackage{textcomp}
\usepackage{color}
\usepackage{array, mathtools}
\usepackage{amsmath,amsfonts,amssymb}
\usepackage{tikz}

\def\benu{\begin{enumerate}}
\def\eenu{\end{enumerate}}
\def\nn{\nonumber}
\def\nn{\nonumber}

\def\l{\left}
\def\r{\right}
\def\d{{\rm d}}
\def\f{\frac}

\def\e{{\rm e}}
\def\k{{\bf k}}
\def\x{{\bf x}}

\begin{document}
\title{Cavity optimization for Unruh effect at small accelerations}
\author{D. Jaffino Stargen}
\email{jaffino@iisermohali.ac.in}
\affiliation{Department of Physical Sciences, IISER Mohali, Knowledge City, 
Sector 81, SAS Nagar, Manauli--140306, Punjab, India}
\author{Kinjalk Lochan}
\email{kinjalk@iisermohali.ac.in}
\affiliation{Department of Physical Sciences, IISER Mohali, Knowledge City, 
Sector 81, SAS Nagar, Manauli--140306, Punjab, India}
\begin{abstract}
One of the primary reasons behind the difficulty in 
observing the Unruh effect is that for 
achievable acceleration scales the finite temperature effects are significant 
only for the low frequency modes of the field. Since the 
density of field modes falls for small frequencies in free space, the  
field modes which are relevant for the thermal effects would be less in number 
to make an observably significant effect.
In this work, we investigate the response of a Unruh-DeWitt detector coupled 
to a massless scalar field which is confined in a long cylindrical cavity. 
The density of field modes inside such a cavity shows a {\it resonance structure} i.e. 
it rises abruptly for some specific cavity configurations.
We show that an accelerating detector inside the cavity exhibits a non-trivial 
excitation and de-excitation rates for {\it small} accelerations 
around such resonance points. If the cavity parameters 
are adjusted to lie in a neighborhood of such resonance points,
the (small) acceleration-induced emission rate can be made much larger than the already 
observable inertial emission rate. We comment on the 
possibilities of employing this detector-field-cavity system in the experimental 
realization of Unruh effect, and argue that the necessity of extremely high 
acceleration can be traded off in favor of precision in cavity manufacturing for realizing 
non-inertial field theoretic effects in laboratory settings. 
\end{abstract}
\maketitle
\noindent
{\it {\bf Introduction}}-- It is well known that the particle content of a 
quantum field is observer dependent \cite{Fulling-1973}, a fact  manifested in 
numerous theoretical arenas, e.g., the Hawking 
radiation, cosmic fluctuations, and Unruh effect 
\cite{Hawking-1974, Davies-1975, Davies-1976, Unruh-1976}. 
In order to estimate the particle content and realize this theoretical
idea, the Unruh-DeWitt detector (UDD) \cite{Unruh-1976, DeWitt-1980} is considered to be an 
operational device.  The UDD is a two-level 
quantum system with the ground state $|E_0\rangle$ and the excited 
state $|E\rangle$, that is moving along a classical worldline  ${\tilde x}(\tau)$, 
where $\tau$ is the proper time in the detector's frame of reference. 
The detector is coupled 
to a quantum field through the interaction Lagrangian 
${\cal L}_{\rm int}[\phi({\tilde x})]=\alpha
m(\tau)\phi[{\tilde x}(\tau)]$, where $\alpha$ is a small coupling constant, 
and $m(\tau)$ is the detector's monopole moment \cite{Unruh-1976, DeWitt-1980} 
which also incorporates a switching function.
In the first-order perturbation theory, the transition probability rate of 
the detector, assuming the scalar field ${\hat \phi}$ 
in its vacuum state $|0\rangle$, is 
given as 
$ {\dot P}(\Delta E) 
 =|\langle E|{\hat m}(0)|E_0\rangle|^2 \times {\dot {\cal F}}(\Delta E)$,
where ${\dot {\cal F}}(\Delta E) =
\int_{-\infty}^{\infty} \d u ~ e^{-i\Delta E u} {\cal W}(u,0)$ is called as 
the response rate of the detector, $\Delta E\equiv E-E_0$, and 
${\cal W}(x,x')\equiv \langle 0|{\hat \phi}(x){\hat \phi}(x')|0\rangle$ is 
the Wightman function of the field. 
The UDD probes 
the vacuum structure of the quantum field through ${\cal W}(x,x')$, and registers 
the excitation of the detector when it absorbs a field quanta. This detector-field 
system has been popularly employed in investigating the effects of quantum fields 
in non-inertial frames, since it encompasses the essential aspects of an atom 
interacting with the electromagnetic field \cite{Martinez-2014}. The response rate 
of a UDD moving in an inertial trajectory can 
be found to be vanishing, since the vacuum structure of the quantum field 
in inertial frames is invariant due to Poincar\'{e} symmetry \cite{Matsas-2008}. 
However, since non-inertial trajectories are not generated by Poincar\'{e} 
transformations, a UDD moving non-inertially detects particles, 
a prime example being -- for uniform acceleration $a$ the detector shows a 
non-vanishing thermal response, known as the Unruh effect 
\cite{Unruh-1976, DeWitt-1980, Matsas-2008}, i.e., 
${\dot {\cal F}}=(\Delta E/2\pi)/(\e^{2\pi\Delta E/a}-1)$. 

Despite being a fundamental prediction, experimental realization of Unruh effect has 
not been made possible due to the demand of extremely high accelerations, 
for appreciable thermal effects one  needs $a \geq 10^{21}~{\rm m/s}^2$
\cite{Matsas-2008}. 
For accelerations small compared to the energy gap $\Delta E$ of the detector,
the response rate is exponentially suppressed, i.e., 
${\dot {\cal F}}\approx (\Delta E/2\pi)\e^{-2\pi\Delta E/a}$. 
This suppression basically originates from the fact that the temperature 
experienced by the accelerating detector is  
vanishingly small for
achievable acceleration scales, since $T\sim \hbar a/{\rm k}_{B}c$. Hence, for such small 
temperatures, the 
significant thermal contribution comes only from the low frequency modes, for which the density 
of field modes (the Bose-Einstein distribution) falls rapidly as 
$\rho(\omega)\sim \omega^2$ in free space, suppressing the response in turn, 
making experimental verification 
of Unruh effect a non-trivial exercise of the current era. 

In response, efforts have been made to enhance the 
detector response for maximum achievable accelerations (in foreseeable future) 
using techniques such as optical cavities \cite{Scully-2003}, 
ultra-intense lasers \cite{Chen-1999, Habs-2008}, and Penning traps \cite{Rogers-1988}. 
Techniques involving capturing the finite temperature effects of an accelerating 
system, such as, monitoring thermal quivering \cite{Raval-1996}, decay of accelerated 
protons \cite{Vanzella-2001}, and radiation emission in Bose-Einstein condensate
\cite{Garay-2000,Retzker-2008} are also proposed. Other than these, 
there are attempts using geometric phases \cite{Mann-2011}, and properly selected 
Fock states \cite{Fuentes-2010} to enhance the effects of non-inertial motion. 
Despite these non-trivial attempts, the efforts are still far  from the 
experimental realization of the Unruh effect (however, see \cite{Kaminer-2021} for a recent claim).

In this letter, we focus on the low acceleration properties of the UDD 
inside an optimized cavity.
To observe Unruh effect for small accelerations, it is important to characterize 
scenarios where the density of field modes is increased appreciably, and the 
correlators of the quantum field are modified 
non-trivially, so that the detector responds in a distinct manner.

The response rate of a UDD moving along a 
given trajectory ${\tilde x}(\tau)$ can be written in a 
more general manner as 
\begin{equation}
{\dot {\cal F}}(\Delta E) \propto 
\int_{0}^{\infty} \d\omega_{k} \rho(\omega_{k}) {\cal I}(\Delta E,\omega_{k})
{\cal J}(\omega_k,\eta^{i}),
\end{equation}
where $\rho(\omega_{k})$ is the density of field modes. The 
quantity ${\cal I}$ depends on the trajectory of the detector through 
field correlations, and determines the field modes which stimulate 
the detector. For example, in the case of 
inertial detector ${\cal I}(\Delta E,\omega_{k})$ is proportional to 
$\delta(\Delta E+\omega_{k})$,  i.e. only modes with energy 
$\omega_{k}=-\Delta E$ can contribute to the response rate of the 
detector, leading to a null response. 
The function ${\cal J}$ depends on the frequency of 
the field modes $\omega_k$, and the coordinates $\eta^{i}$ 
that are held fixed on the trajectory of the detector. Therefore, 
the response rate of the detector 
can be enhanced by the following ways: (i) Increasing the density 
of field modes $\rho(\omega_{k})$ at small $\omega_k$, say, by changing 
the boundary conditions, leading to non-trivial 
changes in the correlators, an aspect missed in the single mode analysis that is
usually employed \cite{Deb-1997,Prants-1999,Scully-2003,Scully-2006,Mann-2011,Lopp-2018}. 
Even for the near resonant 
frequency modes, the response rate for a single mode \cite{Lopp-2018} is suppressed compared 
to the full-mode analysis  (see Supplementary material). 
The analysis in this paper justifiably makes use of the complete set of modes, and not a few modes 
that are near the resonant cavity frequency, which gives an additional enhancement 
channel {\it even at small accelerations}; 
(ii) Choosing the trajectory of the detector appropriately. Even for fixed 
boundary conditions, different non-inertial trajectories associate different quantum 
fluctuations to a given inertial field vacuum \cite{Letaw-1981}, leading to a change in 
${\cal I}(\Delta E,\omega_{k})$ which the detector is sensitive to; 
(iii) Choosing mechanisms, e.g. the stimulated emission, which are extremely 
sensitive to both the boundary conditions and the change in field correlations. 

Making use of these, we demonstrate that for a uniformly accelerated UDD 
in a {\it long} cylindrical cavity, the acceleration-induced 
emission rate can be significantly enhanced, even dominating the inertial spontaneous 
emission, for low accelerations. 
\vskip 5pt
\noindent
{\bf Uniformly accelerating detector in cavity: Role of resonance points}-- 
We consider a UDD inside a cylindrical cavity of radius $R$. The 
length of the cylindrical 
cavity is assumed to be much larger than any scale associated with the detector. 
The scalar field $\phi(x)$ is assumed to satisfy 
Dirichlet boundary condition i.e., $\phi[\rho=R,\theta,z]=0$ in the cylindrical 
polar coordinates. The Wightman
function corresponding to the scalar field inside the cavity can be expressed as 
 \begin{eqnarray}
 \label{eqn:Wightman}
 {\cal W}(x,x')&=&\f{1}{(2\pi R)^2}
 \sum_{m=-\infty}^{\infty} \sum_{n=1}^{\infty}
 \f{J_{m}(\xi_{mn} \rho/R)J_{m}(\xi_{mn} \rho'/R)}{J^2_{|m|+1}(\xi_{mn})} \nn \\
 &\times&\int_{-\infty}^{\infty} \f{\d k_z}{\omega_k}
 \e^{-i\omega_k (t-t'-i\epsilon)} \e^{im(\theta-\theta')} \e^{ik_z(z-z')},
\end{eqnarray}
where $\xi_{mn}$ denotes $n^{{\rm th}}$ zero of the Bessel function $J_m(z)$, 
and $\omega_{k}^2=k_{z}^2+(\xi_{mn}/R)^2$ (see Supplementary material).

For a UDD on a uniformly accelerating trajectory, i.e.,
${\tilde x}(\tau)=[t,\rho,\theta,z]
=(a^{-1}{\rm sinh}a\tau,\rho_0,\theta_0,a^{-1}{\rm cosh}a\tau)$,
where $\rho_0$ and $\theta_0$ are constants, and $a$ denotes proper acceleration of 
the detector,the response rate can be found to be 
\footnote{In the $R \to \infty$ limit, the density of field modes 
reduces to $\rho(\omega_{k}) \propto \omega_k^2$, which is the standard 
density of field modes in free space, provided one makes the following 
replacements: $2\pi \sum_{n=1}^{\infty} \to R\int_{0}^{\infty} \d q$ and 
$\xi_{mn}/R \to q$ and the response rate Eq.~\eqref{eqn:ResponseRate-C} reproduces a thermal form.}
\begin{eqnarray}
\label{Eq:AclrtdResponse}
 & & {\dot {\cal F}}(\Delta E)=\f{1}{2\pi} 
 \int_{0}^{\infty} \d \omega_{k} 
 \underbrace{{\scriptstyle \f{8}{a^2 \e^{\pi \Delta E/a}}
 \frac{K_{2i\Delta E/a}(2 \omega_k/a)}{(2 \omega_k/a)}}} 
 _{{\cal I}(\Delta E,\omega_{k})} \nn \\
 &\times& \sum_{m=-\infty}^{\infty} ~ 
 \underbrace{\sum_{n=1}^{\infty} {\scriptstyle \f{(\omega_k/\pi R^2)}{J^2_{|m|+1}(\xi_{mn})}
 \f{\Theta\l(\omega_k-\xi_{mn}/R\r)}{\sqrt{\omega_k^2-(\xi_{mn}/R)^2}}}}
 _{\rho(\omega_{k})}
 \times 
 \underbrace{{\scriptstyle J_{m}^2(\xi_{mn} \rho_0/R)}}
 _{{\cal J}(\rho_0/R)}, ~~~~~
\end{eqnarray}
where $K_{\nu}(z)$ is the modified Bessel function of second kind, and
$\Theta(x)$ is the Heaviside theta function. One can see that the density of 
field modes $\rho(\omega_k)$ 
has some special features: Firstly, as expected it is independent of the detector parameters -- 
$a$ or $\Delta E$.
Secondly, we can see that $\rho(\omega_k)$ rises abruptly whenever 
$\omega_k^2 \rightarrow (\xi_{mn}/R)^2$, called {\it cavity resonance points},  implying 
the existence of field modes inside the cavity that have very 
large support in terms of 
density of states. How such modes contribute to the 
response rate of the detector is controlled by 
${\cal I}(\Delta E,\omega_{k})$. In order to study that, we further 
evaluate the previous expression to 
\begin{eqnarray}
\label{eqn:ResponseRate-C}
 {\dot {\cal F}}(\Delta E)
 &=& \f{\e^{-\pi \Delta E/a}}{\pi^2 R^2a} 
 \sum_{m=-\infty}^{\infty} \sum_{n=1}^{\infty}
 \f{J_{m}^2(\xi_{mn} \rho_0/R)}{J^2_{|m|+1}(\xi_{mn})} \nn \\
 &\times& K_{i\Delta E/a}^2(\xi_{mn}/Ra).
\end{eqnarray}
In the limit $a\rightarrow 0$ , the function ${\cal I}(\Delta E,\omega_{k})$ 
is proportional to $\delta(\Delta E +\omega_{k})$, as expected (see Supplementary material). 
Thus, in the inertial case 
there  aren't any modes which contribute to the detector response, 
including those at the resonance points. 
However, for the case of
non-inertial detector, the function ${\cal I}(\Delta E,\omega_{k})$ allows for 
the modes around $\omega_k \sim \xi_{mn}/R$ to contribute, with some weightage, 
leading to a non-zero response.

\begin{figure}[!htb]
\begin{center}
\includegraphics[width=8 cm,height=5cm]{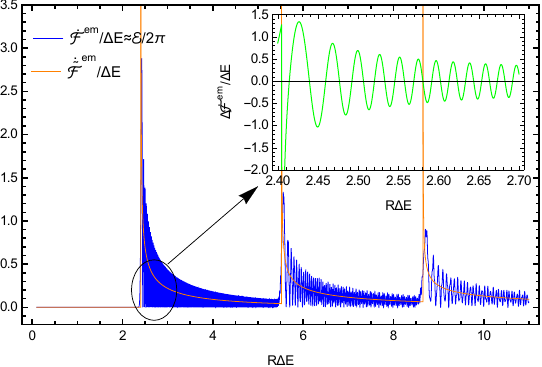} 
\caption{The emission rates for -- the accelerating detector 
${\scriptstyle {\dot {\cal F}}^{\rm em}}$ (which is also proportional to the 
enhancement factor ${\cal E}$ at small accelerations)
and the inertial detector ${\scriptstyle {\dot {\tilde {\cal F}}}^{\rm em}}$
w.r.t. $R\Delta E$, with $\rho_0=0$, and $a/\Delta E=10^{-3}$. 
Inset: The discrete plot for the difference in emission rates of the 
accelerating and the inertial 
detectors $\Delta {\dot {\cal F}}^{\rm em}$ around the first resonance point 
$\xi_{01}$.
The range of $R\Delta E$ and its step size are chosen such that the contribution exactly
at the resonance point $\xi_{01}$ is avoided.} 
\label{fig:De-excitationRate}
\end{center}
\end{figure}

In order to quantify the effects of cavity in enhancing the response 
rate of the accelerating 
detector inside the cavity, when compared to the response rate of an accelerating 
detector in free space ${\dot {\cal F}}_{\cal M}$, we define a 
quantity ${\cal E} \equiv {\dot {\cal F}}/
{\dot {\cal F}}_{\cal M}$, called {\it enhancement} in response rate of the detector.
In the small acceleration limit, i.e., $a\ll \Delta E$, 
we make use of the asymptotic expansion of $K_{i\alpha}(\alpha z)$
for large values of $\alpha$ \cite{Olver:1974}, with 
$\alpha \in \mathbb{R}$ and $|{\rm arg}~z|<\pi$, to approximate 
(see Supplementary material)
\begin{eqnarray}
\label{eqn:EnhancementSmall-ac}
 & & {\cal E}\l(\Delta E\r) \approx 
 \f{4\pi}{(R\Delta E)^2}
 \sum_{m=-\infty}^{\infty} \sum_{n=1}^{\infty}
 \f{J_{m}^2(\xi_{mn} \rho_0/R)}{J^2_{|m|+1}(\xi_{mn})} \\
 &\times&
 \begin{cases} 
 \f{(\beta_{mn}^{<} \Delta E/a)^{1/3}}
 {\l[1-\l(\f{\xi_{mn}}{R\Delta E}\r)^2\r]^{1/2}}~
 {\rm Ai}^2 \l[-(\beta_{mn}^{<}\Delta E/a)^{2/3}\r]; & \f{\xi_{mn}}{R\Delta E} < 1 \\ \\ 
 \f{1}{3^{4/3} \Gamma^2(2/3)} (\Delta E/2a)^{1/3}; & \f{\xi_{mn}}{R\Delta E}=1  \\ \\ 
 \f{(\beta_{mn}^{>} \Delta E/a)^{1/3}}
 {\l[\l(\f{\xi_{mn}}{R\Delta E}\r)^2-1\r]^{1/2}}~
 {\rm Ai}^2\l[(\beta_{mn}^{>} \Delta E/a)^{2/3}\r]; & \f{\xi_{mn}}{R\Delta E} > 1
 \end{cases}, \nn
\end{eqnarray}
where ${\rm Ai}(z)$ is known as the Airy function \cite{Olver:1974}, and
\begin{eqnarray}
 \beta_{mn}^{<} &\equiv& \f{3}{2} {\scriptstyle 
 \l[{\rm sech}^{-1}\l(\f{\xi_{mn}}{R\Delta E}\r)
 -\sqrt{1-\l(\f{\xi_{mn}}{R\Delta E}\r)^2}\r]}, \\ 
 \beta_{mn}^{>} &\equiv& \f{3}{2} {\scriptstyle \l[\sqrt{\l(\f{\xi_{mn}}{R\Delta E}\r)^2-1}
 -{\rm sec}^{-1}\l(\f{\xi_{mn}}{R\Delta E}\r)\r]}. 
\end{eqnarray}
It is evident from Eq.(\ref{eqn:EnhancementSmall-ac}) that in the small acceleration limit the 
enhancement ${\cal E}$ receives a large amplification, 
proportional to $(\Delta E/a)^{1/3}$,
at the resonance points, i.e., $R\Delta E=\xi_{mn}$. 
Thus at small accelerations, if one chooses the 
radius of the cylindrical cavity such that it coincides with one of the 
resonance points, e.g., $R\Delta E=\xi_{01}=2.405$, the enhancement in 
response rate  ${\cal E}$  shows very large amplifications (see Fig.\ref{fig:De-excitationRate}). 

Though the enhancement in detector response diverges at the resonance points 
as $\l(\Delta E/a\r)^{1/3}$ in the limit $a/\Delta E \to 0$, the actual response 
rate of the detector inside the cavity is still small due 
to the exponential suppression of free space response rate for small accelerations, i.e., 
$\lim_{a/\Delta E \to 0} {\dot {\cal F}} 
=\lim_{a/\Delta E \to 0}{\dot {\cal F}}_{\cal M} \times {\cal E} \approx 
(\Delta E/2\pi) \e^{-2\pi \Delta E/a} \times \lim_{a/\Delta E \to 0} {\cal E}.$
It has been argued in \cite{Scully-2003} that the exponential suppression
in the response rate inside a cavity 
can be regulated considerably by introducing non-adiabatic switching
of the detector. Now, if the size of the cylindrical cavity is optimized 
at one of the resonance points in addition to the usage of appropriate switching function, or 
state selection, as proposed in \cite{Scully-2003}, the response rate of the detector can potentially 
be enhanced exponentially. This line of study, however, will be pursued elsewhere. 

In this paper we couple 
the enhancement in response rate ${\cal E}$ at the resonance points, due to the change in 
density of field modes $\rho(\omega_{k})$, to another scheme which is 
extremely sensitive to the change in field correlators, namely the stimulated emission. 
Since stimulated emission is sensitive to the number of particles 
present, and a uniformly 
accelerating detector perceives the Minkowski vacuum as a state with particles, 
one could expect that a uniformly accelerating detector can undergo 
stimulated emission. Higher the number of particles in the Minkowski vacuum 
the detector 
perceives, higher is its emission rate. The emission profile
for a rotating detector was utilized in 
\cite{Lochan-2020} to propose measurable 
detection of non-inertial quantum field theoretic effects. 
In \cite{Kalinski-2005} the emission from a rotating muonic hydrogen atom 
in the so called {\it Trojan states} is shown to be extremely enhanced. 
Thus, modifying the 
density of field modes $\rho(\omega_{k})$ would further strengthen such effects 
which we analyze next.
\vskip 5pt
\noindent
{\bf Acceleration-assisted enhanced emission in cavity: 
Role of ${\cal I}(-\Delta E, \omega_k)$}--
The response rate corresponding to the emission from the UDD can simply be 
obtained as ${\dot {\cal F}}^{\rm em} \l(\Delta E\r)
={\dot {\cal F}} \l(-\Delta E\r)$. 
One can show that the principle of detailed balance is satisfied 
for the detector-field system inside the cavity, i.e.,
${\dot {\cal F}}^{\rm em}/{\dot {\cal F}}=\e^{2\pi \Delta E/a},$ 
leading to a thermal distribution of population
in equilibrium for a collection of such detectors $n_g/n_e = \e^{-2\pi \Delta E/a}$, 
where $n_g$ and $n_e$ denote the number of detectors in the ground and the excited states respectively. 

\begin{figure}[!htb]
\begin{center}
\includegraphics[width=8cm,height=6cm]{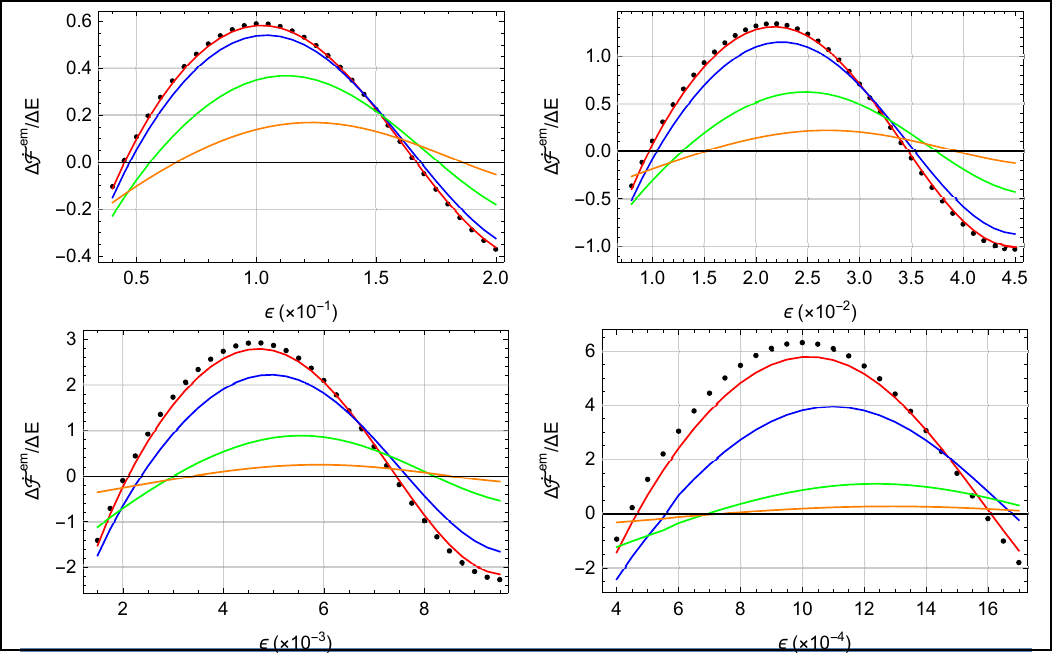} 
\caption{The non-inertial contribution to the emission rate $\Delta {\dot {\cal F}}^{\rm em}$ 
around the first resonance point, 
i.e., $R\Delta E=\xi_{01}+\epsilon$, w.r.t. $\epsilon$ for various values of $a/\Delta E$ 
(plots in the top row represent $a/\Delta E=10^{-2},~10^{-3}$, while for bottom row 
$a/\Delta E=10^{-4},~10^{-5}$). 
The black (dotted), red, blue, green, and orange curves correspond to 
$\beta=0,~10^{-4},~10^{-3},~10^{-2},~10^{-1}$ respectively.}
\label{fig:BandWidth}
\end{center}
\end{figure}

Since only the function ${\cal I}$ in Eq.\eqref{Eq:AclrtdResponse} 
is sensitive to $\Delta E \rightarrow -\Delta E$, the 
emission rate in the cylindrical cavity can be written as
\begin{eqnarray}
\label{eqn:InertialEmissionIntegralForm}
 & & {\dot {\cal F}}^{\rm em}
 \l(\Delta E\r)=\f{1}{\pi R^2} \int_{0}^{\infty} \d \omega_k ~
 {\cal I}(-\Delta E, \omega_k) \nn \\
 &\times& \sum_{m=-\infty}^{\infty} \sum_{n=1}^{\infty}
 \f{J_{m}^2(\xi_{mn} \rho_0/R)}{J^2_{|m|+1}(\xi_{mn})}
 \f{\Theta\l(\omega_k-\xi_{mn}/R\r)}{\sqrt{\omega_k^2-(\xi_{mn}/R)^2}}.
\end{eqnarray}
Note that in the $a\rightarrow 0$ limit ${\cal I}(-\Delta E, \omega_k) \rightarrow\delta(-\Delta E+\omega_{k})$, 
so for an inertial detector only the modes with energy $\omega_{k}=\Delta E$ 
are responsible for the 
emission of the detector. Since the density of field modes 
diverges for modes with energy 
$\omega_{k}=\xi_{mn}/R$, the emission rate becomes divergent if $R \Delta E=\xi_{mn}$, so 
an inertially moving excited detector emits {\it instantaneously} inside such a cavity. 
On the other hand, for uniformly accelerating detector
${\cal I}(-\Delta E, \omega_k) \propto a^{-1}\e^{\pi \Delta E/a} 
K_{-2i\Delta E/a}(2 \omega_k/a)$, there is a 
distribution of modes which determines the emission rate. Some resulting salient 
features are as follows:

\begin{itemize}
\item Since $\delta(-\Delta E+\omega_{k})$ 
in the expression for inertial emission rate 
in Eq.(\ref{eqn:InertialEmissionIntegralForm}) is replaced by a smooth function 
${\cal I}(-\Delta E,\omega_k)$,
the emission rate of the accelerating detector inside a cavity which is optimized at 
its resonant configuration $R \Delta E=\xi_{mn}$ is large, but finite. Thus, if the cavity 
is tuned to be at one of its resonance 
points, while the inertial detector de-excites in no time, the de-excitation of the 
accelerating detector takes finite amount of time, the delay marking the non-inertial effect.

\item Secondly, due to the change in ${\cal I}(-\Delta E, \omega_k)$, caused by 
the accelerated motion, the emission rate of the 
detector in a cavity, optimized {\it slightly away} from the 
resonance points, is larger than that of an inertial detector (see Fig.\ref{fig:De-excitationRate}). 
This is due to the fact that the Delta function (inertial detector) shows a sharper fall 
off away from the resonance points as compared to the smoother function
${\cal I}(-\Delta E, \omega_k)$ of the accelerated detector. 
\end{itemize}
Therefore, in comparison to the inertial detector, acceleration of the detector causes a 
{\it delay} in its emission {\it at the} resonance points of the cavity, but exhibits substantial 
enhancement in emission rate {\it slightly away} from the resonance points. 
Further, in the low acceleration limit the enhancement ${\cal E}$ can be related to the emission 
response rate of the detector ${\dot {\cal F}}^{\rm em}$ as
$\lim_{a/\Delta E \to 0} {\dot {\cal F}}^{\rm em} \approx 
(\Delta E/2\pi) \times \lim_{a/\Delta E \to 0} {\cal E}.$
As the enhancement in response rate ${\cal E}$ of the detector exhibits a 
sharp amplification at the 
resonance points for small accelerations, one could estimate the amount of non-inertial 
contribution in the emission rate of the detector at the resonance points of the cavity.
In order to further 
quantify, we subtract the emission rate of an inertial detector 
${\scriptstyle {\dot {\cal \widetilde F}}^{\rm	em}}$ 
from the non-inertial one, i.e., 
${\scriptstyle \Delta {\dot {\cal F}}^{\rm em}\equiv {\dot {\cal F}}^{\rm em}
-{\dot {\cal \widetilde F}}^{\rm em}}$, obtaining the purely non-inertial contribution 
in the emission rate slightly away from any resonance point as 
\begin{eqnarray}
\label{eqn:Difference}
 & & \Delta {\dot {\cal F}}^{\rm em} \approx 
 \f{2 \Delta E}{(R\Delta E)^2}
 \sum_{m=-\infty}^{\infty} \sum_{n=1}^{\infty}
 \f{J_{m}^2(\xi_{mn} \rho_0/R)}{J^2_{|m|+1}(\xi_{mn})} \\
 &\times&
 \begin{cases} 
 \f{1}{\l[1-\l(\f{\xi_{mn}}{R\Delta E}\r)^2\r]^{1/2}} 
 \{{\scriptstyle (\beta_{mn}^{<} \Delta E/a)^{1/3} 
 {\rm Ai}^2\l[-(\beta_{mn}^{<}\Delta E/a)^{2/3}\r]} \\
 -\f{1}{2\pi}\}; \hskip 4.1cm \f{\xi_{mn}}{R\Delta E} < 1  \\ \\
 \f{(\beta_{mn}^{>} \Delta E/a)^{1/3}}
 {\l[\l(\f{\xi_{mn}}{R\Delta E}\r)^2-1\r]^{1/2}}
 {\scriptstyle {\rm Ai}^2\l[(\beta_{mn}^{>} \Delta E/a)^{2/3}\r]};~~ 
 \f{\xi_{mn}}{R\Delta E} > 1
 \end{cases}. \nn
\end{eqnarray}
Since $ \Delta {\dot {\cal F}}^{\rm em}>0$ amounts to a dominating non-inertial 
emission, we see (Fig.\ref{fig:De-excitationRate}) that the emission rate of 
the accelerating detector can be 
much higher than that of the inertial detector, if the cavity is designed 
to be {\it slightly away from one of its resonance points} i.e. 
$Q_R\equiv 1-(\xi_{mn}/R\Delta E)^2$ is a small (non-zero) number. Since the 
inertial response diverges at the resonance points, very close to the 
resonance points $\Delta {\dot {\cal F}}^{\rm em}$ is a large negative number 
(see Fig.\ref{fig:De-excitationRate} (inset)). However, once one starts moving away 
from the resonance, both inertial and non-inertial emission rates start 
decaying, with the later decaying much slowly in comparison to the inertial delta function. 
As a consequence, closer to the resonance point there is a 
region where the non-inertial response dominates significantly 
(see Fig.\ref{fig:BandWidth}).
{\it Hence, the highly enhanced emission rate of the UDD in a 
slightly off-resonant cavity will clearly be a distinguishable direct realization 
of the Unruh effect.} Thus, the requirement of high acceleration for observing the 
Unruh effect can be compensated for a precise cavity design 
i.e., one with small $Q_R$.
\\ 
{\bf Precision in cavity design}-- 
Since the non-zero acceleration of the detector allows a width of $R\Delta E$ about 
any resonance point (see Fig.\ref{fig:De-excitationRate} (inset)) where the 
non-inertial component dominates, we explore the non-inertial component of the 
emission rate when we go off-resonant by an infinitesimal amount $\epsilon$, 
i.e., $R\Delta E=\xi_{mn}+\epsilon$. As can be seen in Fig.\ref{fig:BandWidth}, even 
for smaller value of acceleration ($ a/\Delta E \sim 10^{-3}$), with increased
precision [$\epsilon \sim (1.5 - 3) \times 10^{-2}$] in cavity design, the emission 
rate of the detector is substantially enhanced.

Moreover, in a realistic experimental setup the cylindrical cavity would 
not be ideal, and the associated $\rho(\omega_k)$  may not really be diverging at 
the resonance points, as discussed above. 
Nevertheless, using a reasonably regularized cavity expressed by
$ \rho(\omega_k)\propto 1/[\beta/R+\sqrt{\omega_{k}^2-(\xi_{mn}/R)^2}]$, 
with the regularization parameter $\beta\ll 1,$  it can be demonstrated that the 
dominance of non-inertial contribution near the resonance points is qualitatively 
independent of the occurrence of divergence in $\rho(\omega_k)$ (see  Fig.\ref{fig:BandWidth}). 
Further, such near-resonance features 
remain present for a realistic cavity with leakage of modes as well, which 
typically provides a Lorentzian broadening for the inertial detectors. In 
the cylindrical cavity this leakage only modifies the function ${\cal I}(\Delta E,\omega_{k})$, 
leaving the structure of density of modes 
$\rho(\omega_k)$, that harbors the resonance, intact. For achievable quality factors 
($\sim 10^{4-6}$) \cite{Bertet-2016,Zmuidzinas-2010}, the modification in the emission for the case of 
Rindler motion is only marginal, i.e., $<1\%$ (see the Supplementary material), 
suggesting the robustness of the scheme.
\vskip 5pt\noindent
{\it {\bf Conclusions}}-- 
To summarize, for small accelerations $a/\Delta E \to 0$, the enhancement ${\cal E}(\Delta E)$ 
in the response rate of the accelerating detector inside a long cylindrical cavity 
diverges as $(\Delta E/a)^{1/3}$ at the resonance points of the 
cavity, i.e., $\xi_{mn}/R\Delta E=1$.  Such resonant configurations of 
cavity can be utilized 
very fruitfully for observing the Unruh effect at small accelerations if one couples 
it with stimulated emission.
Since the emission rate of the inertial detector has a sharp fall off away from the 
resonant frequencies, unlike the accelerating case, to study the non-inertial emission 
rate of the accelerating detector it is advisable to design a cylindrical 
cavity to be in a close neighborhood of a resonance 
point, i.e., $R\Delta E=\xi_{mn}+\epsilon$. In such a cavity, even with 
small enough acceleration, the non-inertial emission rate can be made much larger than the 
inertial emission rate and observable. Similar suppression (dominance) of 
resonant (non-resonant) effects due to the accelerated motion is also observed in 
recent works \cite{Lochan-2020, Kempf-2021}. 

The calculations presented in this paper can easily be 
generalized for other fields, e.g. for a UDD with $\Delta E \sim \text{MHz}$ (e.g. Hydrogen 
atom making a $2p \rightarrow 2s$ transition \cite{Lochan-2020})
inside an optical cavity. The required dimensions of the cavity for such atoms could be 
$(\ell \sim aT^2 \gg a/\Delta E^2 \sim \text{cm}$ 
and $R\sim \xi_{mn}/\Delta E \sim \text{cm})$. For such dimensions even a marginal 
acceleration $a\sim 10^9 ms^{-2}$, that can easily 
be obtained for instance by setting up a thermal gradient \cite{Gallina} of 
$\Delta T = (m a/k_B)\Delta x \sim 1 K$ across the cavity of $\Delta x \sim cm$, leads to a 
significant emission enhancement. Further, 
multiple non-interacting accelerated particles, e.g. a beam of UDDs, can be sent 
inside the cavity, and an integrated enhanced effect can be observed to further strengthen 
the signal \cite{Blencowe-2021}. 

\section*{Acknowledgments}
The authors thank (Late) Prof. T. Padmanabhan and S. Shankaranarayanan for 
useful comments on the manuscript. 
D. J. S thanks the Indian Institute of Science Education and Research
(IISER) Mohali, 
Punjab, India for the financial support. He would also like to acknowledge 
Department of Science and Technology, India, for supporting this work through Project 
No. DST/INSPIRE/04/2016/000571. Research by K. L is partially supported
by the Startup Research Grant of SERB, Government of India (SRG/2019/002202). 
\appendix 
\section{Appendix: Supplementary material}
We consider a Unruh-DeWitt detector, with the ground state $|E_0\rangle$ and the 
excited state $|E\rangle$, coupled to a massless scalar field $\phi(x)$ 
which is confined inside a cylindrical cavity of radius $R$. The confined scalar field 
satisfies Dirichlet boundary condition. The interaction 
between the detector and the scalar field is described by the interaction Lagrangian 
${\cal L}_{\rm int}[\phi({\tilde x}(\tau)]=\alpha m(\tau) \phi({\tilde x}(\tau)$.
If the initial state of the detector-field system is $|0\rangle \otimes |E_0\rangle$, and 
the final state is $|{\bf k}\rangle \otimes |E\rangle$, where $|0\rangle$ and 
$|{\bf k}\rangle$ are respectively the vacuum state and the one-particle state of 
the scalar field, then the transition probability associated with this process in 
the first-order perturbation theory can be written as 
\begin{equation}
 {\cal A}_{\bf k}=i \langle E| \otimes \langle {\bf k}| 
 \int_{-\infty}^{\infty} \d \tau {\hat m}(\tau) {\hat \phi}[{\tilde x}(\tau)] 
 |0\rangle \otimes |E_0\rangle.
\end{equation}
Integrating over all possible one-particle states of the field, one can obtain the 
transition probability as 
\begin{eqnarray}
 P_{E_0 \to E}&=&|\langle E|{\hat m}(0)|E_0\rangle|^2 \\
 &\times& \int_{-\infty}^{\infty} 
 \int_{-\infty} ^{\infty} \d \tau \d \tau' \e^{-i\Delta E (\tau-\tau')} 
 {\cal W}[x(\tau),x(\tau')], \nn 
\end{eqnarray}
where ${\cal W}(x,x')=\langle 0| {\hat \phi}(x) {\hat \phi}(x.) |0\rangle$ is the 
Wightman function corresponding to the field. 

If the detector is allowed to move along the integral curve of a Killing vector field, 
then the transition probability can be reduced to transition probability rate as 
\begin{eqnarray}
 {\dot P}(\Delta E) \equiv \lim_{v \to \infty} \f{P_{E_0 \to E}}{v}
 =|\langle E|{\hat m}(0)|E_0\rangle|^2 \times {\dot {\cal F}}(\Delta E),
\end{eqnarray}
where ${\dot {\cal F}}$ is called as response rate of the detector, and is 
\begin{eqnarray}
 {\dot {\cal F}}(\Delta E)=\int_{-\infty}^{\infty} \d u \e^{-i\Delta Eu} 
 {\cal W}(u,0),
\end{eqnarray}
with $u=\tau-\tau'$ and $v=(\tau+\tau')/2$. Note that the response rate 
${\dot {\cal F}}$ of the detector is just the Fourier transform of the 
pullback of the Wightman function on the trajectory of the detector.

The scalar field ${\hat \phi}$ confined inside the cavity 
satisfying Dirichlet boundary condition is 
\begin{eqnarray}
 {\hat \phi}(x)&=&\f{1}{2\pi R}\sum_{m=-\infty}^{\infty}\sum_{n=1}^{\infty}
 \f{J_{m}(\xi_{mn} r/R)}{J_{|m|+1}(\xi_{mn})} 
 \int_{-\infty}^{\infty} \f{\d k_z}{\sqrt{\omega_k}} \\ 
 &\times& 
 \l({\hat a}_{\bf k} \e^{-i\omega_k t} \e^{im\theta} \e^{ik_z z}
 +{\hat a}^{\dagger}_{\bf k} \e^{i\omega_k t} \e^{-im\theta} \e^{-ik_z z}\r), \nn
\end{eqnarray}
where $[{\hat a}_{\bf k},{\hat a}^{\dagger}_{\bf k'}]
=\delta_{mm'} \delta_{nn'} \delta(k_{z}-k'_{z})$,
$[{\hat a}_{\bf k},{\hat a}_{\bf k'}]=0$, and $[{\hat a}^{\dagger}_{\bf k},
{\hat a}^{\dagger}_{\bf k'}]=0$.
Making use of the expression for the field ${\hat \phi}$ inside the cavity, we find the 
Wightman function as 
\begin{eqnarray}
 {\cal W}(x,x')&=&\f{1}{(2\pi R)^2}
 \sum_{m=-\infty}^{\infty} \sum_{n=1}^{\infty}
 \f{J_{m}(\xi_{mn} \rho/R)J_{m}(\xi_{mn} \rho'/R)}{J^2_{|m|+1}(\xi_{mn})} \nn \\
 &\times& \int_{-\infty}^{\infty} \f{\d k_z}{\omega_k}
 \e^{-i\omega_k (t-t'-i\epsilon)} \e^{im(\theta-\theta')} \e^{ik_z(z-z')}, 
\end{eqnarray}
where $\omega_k^2=(\xi_{mn}/R)^2+k_z^2$.

Substituting the trajectory of an accelerating detector
$[t=a^{-1} {\rm sinh}(a\tau),\rho=\rho_0,\theta=\theta_0,
z=a^{-1} {\rm cosh}(a\tau)]$ in the expression for 
Wightman function, and using it in the expression for response rate 
${\dot {\cal F}}$ of the detector, we find
\begin{eqnarray}
\label{Appendix:1}
 {\dot {\cal F}}(\Delta E)
 &=& \f{2}{(2\pi R)^2} \sum_{m=-\infty}^{\infty} \sum_{n=1}^{\infty}
 \f{J_{m}^2(\xi_{mn} \rho_0/R)}{J^2_{|m|+1}(\xi_{mn})} \\
 &\times& \int_{0}^{\infty} \d \omega_k 
 \f{\Theta(\omega_k-\xi_{mn}/R)}{\sqrt{\omega_{k}^2-(\xi_{mn}/R)^2}} \nn \\
 &\times& \int_{-\infty}^{\infty} \d u e^{-i\Delta Eu}
 {\rm exp}\l\{-2i(\omega_k/a) {\rm sinh}(au/2)\r\}. \nn
\end{eqnarray}
In the limit $a\to 0$ in Eq.(\ref{Appendix:1}), we arrive at the 
response rate of the inertial detector inside the cavity as
\begin{eqnarray}
 & & {\dot {\cal F}}(\Delta E)
 =\f{1}{\pi R^2} \sum_{m=-\infty}^{\infty} \sum_{n=1}^{\infty}
 \f{J_{m}^2(\xi_{mn} \rho_0/R)}{J^2_{|m|+1}(\xi_{mn})} \nn \\
 &\times& \int_{0}^{\infty} \d \omega_k 
 \f{\Theta(\omega_k-\xi_{mn}/R)}{\sqrt{\omega_{k}^2-(\xi_{mn}/R)^2}}~ 
 \delta(\Delta E+\omega_k),
\end{eqnarray}
which is vanishing since the argument of the delta function is positive 
throughout the range of $\omega_k$.

Evaluating the integrals in the expression for response rate of the 
accelerating detector in Eq.(\ref{Appendix:1}), we obtain 
\begin{eqnarray}
\label{Appendix:2}
 {\dot {\cal F}}(\Delta E)&=&\f{1}{\pi^2 R} \f{\e^{-\pi\Delta E/a}}{Ra} 
 \sum_{m=-\infty}^{\infty} \sum_{n=1}^{\infty}
 \f{J_{m}^2(\xi_{mn} \rho_0/R)}{J^2_{|m|+1}(\xi_{mn})} \nn \\
 &\times& K_{i\Delta E/a}^2(\xi_{mn}/Ra). 
\end{eqnarray}
Making use of the asymptotic expansion of $K_{i\alpha}(\alpha x)$ 
for large positive values of $\alpha$ \cite{Olver:1974}, which is 
\begin{eqnarray}
{\scriptstyle 
K_{i\alpha}(\alpha x)=\f{\e^{-\pi \alpha/2}}{\sqrt{\alpha}} \pi \sqrt{2}}
 \begin{cases}
 {\scriptstyle \f{(\beta^{<} \alpha)^{1/6}}
 {\l(1-x^2\r)^{1/4}}~
 {\rm Ai}\left[-(\beta^{<} \alpha)^{2/3}\right];} & {\scriptstyle x<1}  \\ \\
 {\scriptstyle \f{\alpha^{1/6}}{3^{2/3} \Gamma(2/3)}}; & {\scriptstyle x=1} \\ \\
 {\scriptstyle \f{(\beta^{>} \alpha)^{1/6}}
 {\l(x^2-1\r)^{1/4}}~
 {\rm Ai}\left[(\beta^{>} \alpha)^{2/3}\right];} & {\scriptstyle x>1}
 \end{cases}, 
\end{eqnarray}
where ${\rm Ai}(z)$ is the Airy function, with 
\begin{eqnarray}
 \beta^{<} &\equiv& \f{3}{2} \l({\rm sech}^{-1}x
 -\sqrt{1-x^2}\r), \\ 
 \beta^{>} &\equiv& \f{3}{2} \l(\sqrt{x^2-1}
 -{\rm sec}^{-1}x\r),
\end{eqnarray}
in Eq.(\ref{Appendix:2}), we obtain 
\begin{eqnarray}
 & & {\dot {\cal F}}(\Delta E) \approx 
 \f{2}{R} \f{\e^{-2\pi\Delta E/a}}{R\Delta E}
 \sum_{m=-\infty}^{\infty} \sum_{n=1}^{\infty}
 \f{J_{m}^2(\xi_{mn} \rho_0/R)}{J^2_{|m|+1}(\xi_{mn})} \\
 &\times&
 \begin{cases} 
 \f{(\beta_{mn}^{<} \Delta E/a)^{1/3}}
 {\left[1-\l(\f{\xi_{mn}}{R\Delta E}\r)^2\right]^{1/2}}~
 {\rm Ai}^2 \left[-(\beta_{mn}^{<}\Delta E/a)^{2/3}\right]; & \f{\xi_{mn}}{R\Delta E} < 1 \\ \\ 
 \f{1}{3^{4/3} \Gamma^2(2/3)} (\Delta E/2a)^{1/3}; & \f{\xi_{mn}}{R\Delta E}=1  \\ \\ 
 \f{(\beta_{mn}^{>} \Delta E/a)^{1/3}}
 {\left[\l(\f{\xi_{mn}}{R\Delta E}\r)^2-1\right]^{1/2}}~
 {\rm Ai}^2\left[(\beta_{mn}^{>} \Delta E/a)^{2/3}\right]; & \f{\xi_{mn}}{R\Delta E} > 1
 \end{cases}, \nn
\end{eqnarray}
where 
\begin{eqnarray}
 \beta_{mn}^{<} &\equiv& \f{3}{2} {\scriptstyle 
 \left[{\rm sech}^{-1}\l(\f{\xi_{mn}}{R\Delta E}\r)
 -\sqrt{1-\l(\f{\xi_{mn}}{R\Delta E}\r)^2}\right]}, \nn  \\ 
 \beta_{mn}^{>} &\equiv& \f{3}{2} {\scriptstyle \left[\sqrt{\l(\f{\xi_{mn}}{R\Delta E}\r)^2-1}
 -{\rm sec}^{-1}\l(\f{\xi_{mn}}{R\Delta E}\r)\right]}. \nn
\end{eqnarray}
\section{Contribution due to a single-mode}
When it comes to investigating the interaction of atoms with quantum fields confined inside 
cavities, it is quite common to employ the 
single-mode approximation where the field inside the cavity is assumed to be the 
field modes that are in resonance with the resonant configuration of the cavity 
\cite{Deb-1997, Prants-1999, Scully-2003, Mann-2011, Lopp-2018}. 
Recently it was shown that the single-mode approximation becomes inaccurate when it comes to 
trajectories of detectors/atoms that are relativistic and non-inertial \cite{Lopp-2018}. 
The enhancement discussed in this paper at low accelerations around the resonant configurations 
of the cavity is due to the divergence in the density of field modes $\rho(\omega_{k})$, which 
cannot be captured by the single-mode analysis. 
To see this, we write the scalar field inside a cylindrical cavity of finite length 
$L$ with boundary conditions 
\begin{eqnarray}
 \phi(\rho=R,\phi,z)&=&0, \\
 \phi(\rho,\phi,z=\pm L/2)&=&0,
\end{eqnarray}
as 
\begin{eqnarray}
 &&{\hat \phi}(t,\x)=\f{1}{\sqrt{\pi R^2 L}} 
 \sum_{l=-\infty}^{\infty} \sum_{m=-\infty}^{\infty}\sum_{n=1}^{\infty} 
 \f{J_{m}(\xi_{mn} \rho/R)}{J_{|m|+1}(\xi_{mn})} \nn \\ 
 &\times& \Biggl\{{\hat a}_{k} 
 \Biggl[\f{1}{\sqrt{2\omega^{(s)}_{k}}} {\rm cos}\l(\f{(2l+1)\pi}{L}z\r) \e^{-i\omega^{(s)}_k t} \nn \\  
 &+&\f{i}{\sqrt{2\omega^{(a)}_{k}}} {\rm sin}\l(\f{2l\pi}{L}z\r) 
 \e^{-i\omega^{(a)}_k t}\Biggr]  \e^{im\theta} \nn \\ 
 &+&{\hat a}^{\dagger}_{k} 
 \Biggl[\f{1}{\sqrt{2\omega^{(s)}_{k}}} {\rm cos}\l(\f{(2l+1)\pi}{L}z\r) 
 \e^{i\omega^{(s)}_k t} \nn \\ 
 &-&\f{i}{\sqrt{2\omega^{(a)}_{k}}} {\rm sin}\l(\f{2l\pi}{L}z\r) 
 \e^{i\omega^{(a)}_k t}\Biggr] \e^{-im\theta}\Biggr\}, 
\end{eqnarray}
with $k\equiv(l,m,n)$, 
$\omega^{(s)}_{k}=\sqrt{\l(\f{\xi_{mn}}{R}\r)^2+\l(\f{(2l+1)\pi}{L}\r)^2}$, 
$\omega^{(a)}_{k}=\sqrt{\l(\f{\xi_{mn}}{R}\r)^2+\l(\f{2l\pi}{L}\r)^2}$, and 
\begin{eqnarray}
 && [{\hat a}_{k},{\hat a}_{k'}^{\dagger}]=\delta_{ll'} \delta_{mm'} \delta_{nn'}, \\ 
 && [{\hat a}_{k},{\hat a}_{k'}]=0, ~~ 
 [{\hat a}_{k}^{\dagger},{\hat a}_{k'}^{\dagger}]=0. 
\end{eqnarray}
The Wightman function corresponding to the field confined inside the cavity can be found 
to be 
\begin{eqnarray}
 &&{\cal W}(x,x')=\f{1}{\pi R^2L} 
 \sum_{l=-\infty}^{\infty} \sum_{m=-\infty}^{\infty}\sum_{n=1}^{\infty} \nn \\ 
 &\times&\f{J_{m}(\xi_{mn} \rho/R)}{J_{|m|+1}(\xi_{mn})} 
 \f{J_{m}(\xi_{mn} \rho'/R)}{J_{|m|+1}(\xi_{mn})} \e^{im(\theta-\theta')} \nn \\ 
 &\times& \Biggl\{\f{1}{2\omega^{(s)}_{k}} {\rm cos}\l(\f{(2l+1)\pi}{L}z\r) 
 {\rm cos}\l(\f{(2l+1)\pi}{L}z'\r) \nn \\ 
 &\times& \e^{-i\omega^{(s)}_k (t-t')} \nn \\ 
 &-&\f{i}{2\sqrt{\omega^{(s)}_{k} \omega^{(a)}_{k}}} {\rm cos}\l(\f{(2l+1)\pi}{L}z\r) 
 {\rm sin}\l(\f{2l\pi}{L}z'\r) \nn \\ 
 &\times& \e^{-i\omega^{(s)}_kt} \e^{i\omega^{(a)}_{k}t'} \nn \\ 
 &+&\f{i}{2\sqrt{\omega^{(s)}_{k} \omega^{(a)}_{k}}} {\rm sin}\l(\f{2l\pi}{L}z\r) 
 {\rm cos}\l(\f{(2l+1)\pi}{L}z'\r) \nn \\ 
 &\times& \e^{i\omega^{(s)}_kt'} \e^{-i\omega^{(a)}_{k}t} \nn \\ 
 &+&\f{1}{2\omega^{(a)}_{k}} {\rm sin}\l(\f{2l\pi}{L}z\r) {\rm sin}\l(\f{2l\pi}{L}z'\r) \nn \\ 
 &\times& \e^{-i\omega^{(a)}_k (t-t')}\Biggr\}. 
\end{eqnarray}
\begin{figure}[!htb]
\begin{center}
\includegraphics[width=8 cm,height=5cm]{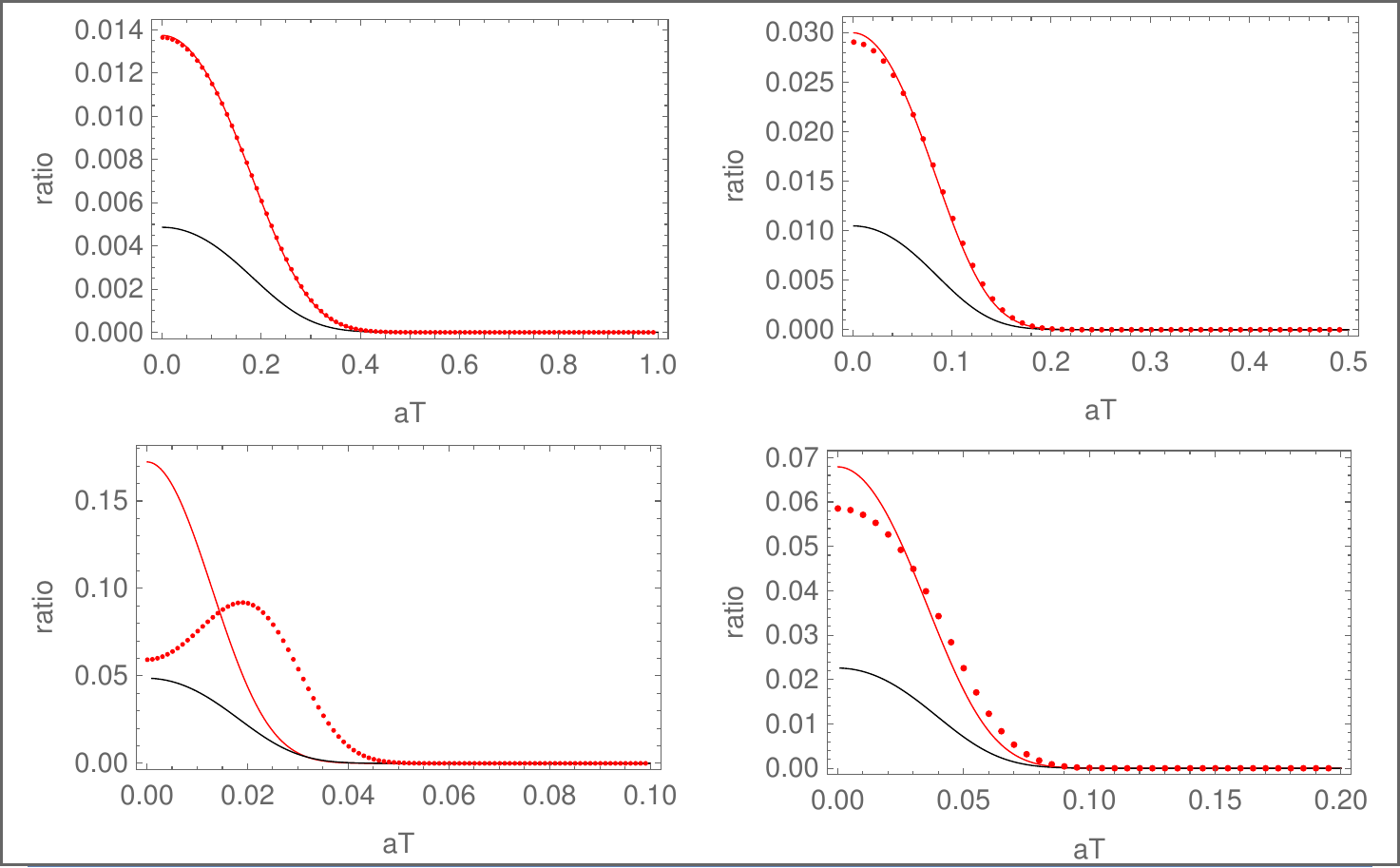} 
\caption{The ratio $r$ of emission rates due to a (symmetric) single-mode 
($l'=0,~m'=0,~n'=1$) and complete set of modes around the first resonance 
point, i.e., $R\Delta E=\xi_{01}$, is plotted with respect to. $aT$ for various values of 
$a/\Delta E$ (clockwise from left top, $a/\Delta E=10^{-2},~10^{-3},~10^{-4},~10^{-5}$). 
The dotted red, black, and joined red curves correspond to 
$R\Delta E-\xi_{01}=-10^{-3},0,10^{-3}$ respectively. Note that we 
have assumed $L/R=10^{3}$ for all the plots.}
\label{fig:ratio}
\end{center}
\end{figure}
If one assumes the length $L$ of the cavity is much 
larger that the length scales involved in the system, i.e., $a^{-1}$, $\Delta E^{-1}$, 
and $R$, then the finite-time ($T$) response rate of the uniformly accelerating detector can be evaluated as 
\begin{eqnarray}
  &&\dot {\cal F}\l(\Delta E\r)\approx\f{\e^{-\pi\Delta E/a}}{\pi (aR) RL} 
 \sum_{l=-\infty}^{\infty} \sum_{m=-\infty}^{\infty}\sum_{n=1}^{\infty} 
 \f{J^2_{m}(\xi_{mn} \rho_0/R)}{J^2_{|m|+1}(\xi_{mn})} \nn \\ 
 &\times& \Biggl\{\Biggl(\f{1}{\omega^{(s)}_{k}} 
 {\rm cos}\left[\f{\Delta E}{a} 
 \ln\l(\f{L\omega^{(s)}_{k}-(2l+1)\pi}{L\omega^{(s)}_{k}+(2l+1)\pi}\r)\right] \nn \\ 
 &-&\f{1}{\omega^{(a)}_{k}} 
 {\rm cos}\left[\f{\Delta E}{a} \ln\l(\f{L\omega^{(a)}_{k}-2l\pi}{L\omega^{(a)}_{k}+2l\pi}\r)\right] 
  \Biggr) \nn \\ 
 &\times& K_{2i\Delta E/a}\l(\f{2\xi_{mn}}{Ra}{\rm cosh}aT\r) \nn \\ 
 &+&\f{1}{\omega^{(s)}_{k}} 
 K_{2i\Delta E/a}\left[\f{2}{a}\l(\omega^{(s)}_{k}{\rm cosh}aT-\f{(2l+1)\pi}{L}{\rm sinh}aT\r)\right] \nn \\ 
 &+&\f{1}{\omega^{(a)}_{k}} 
 K_{2i\Delta E/a} 
 \left[\f{2}{a}\l(\omega^{(a)}_{k}{\rm cosh}aT-\f{2l\pi}{L}{\rm sinh}aT\r)\right]\Biggr\}. 
\end{eqnarray}

Choosing one particular (symmetric) mode $\omega^{(s)}_{k'}$ we calculate the response rate 
${\dot {\cal F}}^{(s)}_{\rm sm}$ 
of the detector due to this single-mode to be 
\begin{eqnarray}
  &&{\dot {\cal F}}^{(s)}_{\rm sm}\l(\Delta E\r)\approx\f{\e^{-\pi\Delta E/a}}{\pi (aR) RL} 
 \f{J^2_{m'}(\xi_{m'n'} \rho_0/R)}{J^2_{|m'|+1}(\xi_{m'n'})} \\ 
 &\times& \Biggl\{\f{1}{\omega^{(s)}_{k'}} 
 {\rm cos}\left[\f{\Delta E}{a} 
 \ln\l(\f{L\omega^{(s)}_{k'}-(2l'+1)\pi}{L\omega^{(s)}_{k'}+(2l'+1)\pi}\r)\right] \nn \\ 
 &\times& K_{2i\Delta E/a}\l(\f{2\xi_{m'n'}}{Ra}{\rm cosh}aT\r) \nn \\ 
 &+&\f{1}{\omega^{(s)}_{k'}} 
 K_{2i\Delta E/a}
 \left[\f{2}{a}\l(\omega^{(s)}_{k'}{\rm cosh}aT-\f{(2l'+1)\pi}{L}{\rm sinh}aT\r)\right]\Biggr\}. \nn 
\end{eqnarray}

Using this 
we calculate the ratio $r\equiv{\dot {\cal F}}^{(s)}_{\rm sm}/{\dot {\cal F}}$ to 
compare the response rate due to the single-mode viz-a-viz the same 
due to the complete set of modes, and plot it with respect to $aT$ as 
shown in Fig.(\ref{fig:ratio}). It is evident from Fig.(\ref{fig:ratio}) that 
the contribution due to a single-mode is insignificant in the late-time limit 
($T\to\infty$) against the contribution due to the complete set of field modes \cite{Lopp-2018}. 
Also, the single-mode approximation evidently misses the 
near-resonance ($R\Delta E \approx \xi_{mn}$) 
enhancement in the response rate brought in by the complete set of field modes, 
particularly through the density of field modes $\rho(\omega_{k})$. 
\section{Cavity with Loss}
For realistic cavities with non-zero leakage, assuming the field encompasses the information 
about dissipation, one could write 
\begin{eqnarray}
 &&{\hat \phi}(t,\x)=\f{1}{2\pi R}\sum_{m=-\infty}^{\infty}\sum_{n=1}^{\infty} \e^{-\alpha_{mn}|t|} 
 \f{J_{m}(\xi_{mn} r/R)}{J_{|m|+1}(\xi_{mn})} \\ 
 &\times&\int_{-\infty}^{\infty} \f{\d k_z}{\sqrt{\omega_k}}  
 \l({\hat a}_{\k} \e^{-i\omega_k t} \e^{im\theta} \e^{ik_z z}
 +{\hat a}^{\dagger}_{\k} \e^{i\omega_k t} \e^{-im\theta} \e^{-ik_z z}\r), \nn 
\end{eqnarray}
where $\alpha_{mn}= \xi_{mn}/R Q$ are the inertial leakage factors of the modes 
inside the cavity, and they depend on the 
quality factor $Q$ of the cavity. Making use of this in the two point function, 
we obtain the 
expression for the emission probability of the detector as 
\begin{eqnarray}
{\cal P}(\Delta E) &=& \int_{-\infty}^{\infty} \d v \int_{-\infty}^{\infty} \d u  e^{i\Delta E  u} 
 \langle0| {\hat \phi}(\tau){\hat \phi}(\tau')|0\rangle \nn \\ 
 &\equiv&\int_0^{\infty}  \d v~{\dot {\cal F}}^{\text{em}}_Q(\Delta E), 
\end{eqnarray}
where $u=\tau-\tau'$ and $v=(\tau+\tau')/2$. 
The emission response rate of an inertial detector, i.e., the trajectory of the detector be 
$(\gamma \tau,\rho_0,\theta_0,\gamma v \tau)$, 
can be found to be $v$ dependent as 
\begin{eqnarray}
 &&{\dot {\cal F}}^{\text{em}}_{Q}\l(\Delta E\r)=\int_{0}^{\infty} \d \omega_{k} \nn \\ 
 &\times& \sum_{m=-\infty}^{\infty} ~ 
 \sum_{n=1}^{\infty} {\scriptstyle \f{(\omega_k/\pi R^2)}{J^2_{|m|+1}(\xi_{mn})}
 \f{\Theta\l(\omega_k-\xi_{mn}/R\r)}{\sqrt{\omega_k^2-(\xi_{mn}/R)^2}}} \nn \\
 &\times&{\scriptstyle \e^{-2\gamma\alpha_{mn} v} 
 \Biggl\{\l[1-\f{(\omega_k-\Delta E)^2}{[(\gamma\alpha_{mn})^2+(\omega_k-\Delta E)^2]}\r] 
 \f{{\rm sin}[2(\omega_k-\Delta E)v]}{(\omega_k-\Delta E)}} \\ 
 &+&{\scriptstyle \f{\gamma\alpha_{mn}}{[(\gamma\alpha_{mn})^2+(\omega_k-\Delta E)^2]} 
 ~{\rm cos}[2(\omega_k-\Delta E)v]\Biggr\} \times \f{J_{m}^2(\xi_{mn} \rho_0/R)}{\omega_k}}. \nn 
\end{eqnarray}
In the zero loss limit ($\{\alpha_{mn}\} \to 0$) we obtain the standard inertial expression for the emission rate as 
\begin{eqnarray}
 &&{\dot {\cal F}}^{\text{em}}_{Q\to \infty}(\Delta E)= 
 \f{1}{\pi R^2} \sum_{m=-\infty}^{\infty} \sum_{n=1}^{\infty}
 \f{J_{m}^2(\xi_{mn} \rho_0/R)}{J^2_{|m|+1}(\xi_{mn})} \nn \\  
 &\times&\int_{0}^{\infty} \d \omega_k \f{\Theta\l(\omega_k-\xi_{mn}/R\r)}{\sqrt{\omega_k^2-(\xi_{mn}/R)^2}} 
 \delta(\omega_k-\Delta E). 
\end{eqnarray}
In the limit $v\to 0$, the rate has a Lorentzian support for various modes
\begin{eqnarray}
 &&\lim_{v\to 0} {\dot {\cal F}}^{\text{em}}_Q(\Delta E)= 
 \int_{0}^{\infty} \d \omega_{k} \nn \\
 &\times&\sum_{m=-\infty}^{\infty} ~ 
 \underbrace{\sum_{n=1}^{\infty} {\scriptstyle \f{(\omega_k/\pi R^2)}{J^2_{|m|+1}(\xi_{mn})}
 \f{\Theta\l(\omega_k-\xi_{mn}/R\r)}{\sqrt{\omega_k^2-(\xi_{mn}/R)^2}}}}_{\rho(\omega_{k})} \nn \\ 
 &&\underbrace{{\scriptstyle \f{\gamma\alpha_{mn}}{\omega_k[(\gamma\alpha_{mn})^2 
  +(\omega_k-\Delta E)^2]}}}_{{\cal I}(-\Delta E,\omega_{k})} \times 
 \underbrace{{\scriptstyle J_{m}^2(\xi_{mn} \rho_0/R)}}
 _{{\cal J}(\rho_0/R)}. 
\end{eqnarray}
Similarly, the emission rate of the detector in a non-inertial Rindler trajectory, i.e., 
$[a^{-1} {\rm sinh}(a\tau),\rho_0,\theta_0,a^{-1} {\rm cosh}(a\tau)]$ in the limit $av\to0$ can be obtained as 
\begin{eqnarray}
 &&\lim_{av \to 0} {\dot {\cal F}}^{\text{em}}_Q(\Delta E)=\int_{0}^{\infty} \d \omega_{k}  \nn \\ 
 &\times&\sum_{m=-\infty}^{\infty} 
 \underbrace{\sum_{n=1}^{\infty} {\scriptstyle \f{(\omega_k/\pi R^2)}{J^2_{|m|+1}(\xi_{mn})} 
 \f{\Theta\l(\omega_k-\xi_{mn}/R\r)}{\sqrt{\omega_k^2-(\xi_{mn}/R)^2}}}}_{\rho(\omega_{k})} 
 \times\underbrace{{\scriptstyle J_{m}^2(\xi_{mn} \rho_0/R)}}_{{\cal J}(\rho_0/R)} \nn \\ 
 &\times&\underbrace{{\scriptstyle \frac{1}{\omega_k} 
 \text{Re}\Biggl(\int_{0}^{\infty} du~e^{i\Delta E u} 
 {\rm exp}\left[-i\f{2(\omega_k +\alpha_{mn})}{a} {\rm sinh}(au/2)\right]\Biggr)}}_{{\cal I}(-\Delta E,\omega_{k})}, \\ 
 &&=\f{1}{\pi^2 R} \f{(2\Delta E/a)}{(R\Delta E)} \sum_{m=-\infty}^{\infty} \sum_{n=1}^{\infty}
 \f{J_{m}^2(\xi_{mn} \rho_0/R)}{J^2_{|m|+1}(\xi_{mn})} \nn \\
 &\times&\text{Re}\biggl\{\int_{0}^{\infty} dx~e^{-(2\Delta E/a)\l[(R\alpha_{mn}/R\Delta E) \sinh x-ix\r]} \nn \\ 
 &\times&K_0\l[i(2\Delta E/a)\f{\xi_{mn}}{R\Delta E} {\rm sinh}x\r]\biggr\}, \label{UnruhCavityLoss} 
\end{eqnarray}
which for the no loss case $(\{\alpha_{mn}\}=0)$ reproduces Eq.\ref{eqn:ResponseRate-C}. 
In the limit $av\to0$, 
the departure in emission rate from the same for an ideal cavity (no loss) for various values of the quality factor 
can be computed by evaluating the 
integral in Eq.\eqref{UnruhCavityLoss}. 
\begin{table} [h!]
\centering
\begin{tabular}{|l|l|l|}
\hline
 $Q$ & ${\dot {\cal F}}^{\text{em}}_Q/\Delta E$ & 
 $(1-{\dot {\cal F}}^{\text{em}}_Q/{\dot {\cal F}}^{\text{em}}_{Q\to\infty})\times100 \%$ \\ \hline
 $10^3$ &$\sim 1.226 $ & $~~~~~~~\sim 3\%$ \\ \hline
 $10^4$ &$\sim 1.259 $ & $~~~~~~~\sim 0.3\%$\\ \hline 
 $10^5$ &$\sim 1.262 $ & $~~~~~~~\sim 0.03\%$\\ \hline 
 $10^6$ &$\sim 1.263 $ & $~~~~~~~\sim 0.003\%$ \\ 
\hline   
\end{tabular}
\caption{The $Q$-modified emission rate and the $\%$ deviation of the same from the lossless ($Q\to\infty$) 
emission rate for different quality factors.}
\label{table:1}
\end{table}
An estimate for the emission rate
in a realistic cavity set up 
$(\Delta E \sim  \text{MHz}; a \sim 10^{12} ms^{-2};\rho_0=0; R\Delta E =\xi_{01} + 0.1 $) 
is presented in Table \ref{table:1}.

\end{document}